\documentclass[journal]{IEEEtran}
\usepackage{amsmath}
\pdfoutput=1
\usepackage{hyperref} 
\usepackage{cleveref}
\usepackage{amsfonts}
\usepackage{amssymb}
\usepackage{graphicx,times} 
\usepackage{subfigure} 
\usepackage{CJK}         
\usepackage{cite}
\usepackage{siunitx}
\ifCLASSINFOpdf
\else
   \usepackage[dvips]{graphicx}
\fi
\usepackage{url}

\usepackage{graphicx}

\begin{document}

\title{Deep Neural Network-Based Quantized Signal Reconstruction for DOA Estimation}

\author{Weifeng Han, Peng Chen, \IEEEmembership{Member, IEEE}, Zhenxin Cao, \IEEEmembership{Member, IEEE}
\thanks{The authors are with the State Key Laboratory of Millimeter Waves, Southeast University, email: \{hanweifeng, chenpengseu, caozx\}@seu.edu.cn.}}

\markboth{Journal of \LaTeX\ Class Files, Vol. 14, No. 8, August 2015}
{Shell \MakeLowercase{\textit{et al.}}: Bare Demo of IEEEtran.cls for IEEE Journals}
\maketitle

\begin{abstract}
	
For a massive multiple-input-multiple-output (MIMO) system using intelligent reflecting surface (IRS) equipped with radio frequency (RF) chains, the multi-channel RF chains are expensive compared to passive IRS, especially, when the high-resolution and high-speed analog to digital converters (ADC) are used in each RF channel. In this letter, a direction of angle (DOA) estimation problem is investigated with low-cost ADC in IRS, and we propose a deep neural network (DNN) as a recovery method for the low-resolution sampled signal. Different from the existing denoising convolutional neural network (DnCNN) for Gaussian noise, the proposed DNN with fully connected (FC) layers estimates the quantization noise caused by the ADC. Then, the denoised signal is subjected to the DOA estimation, and the recovery performance for the quantized signal is evaluated by  DOA estimation. Simulation results show that under the same training conditions, the better reconstruction performance is achieved by the proposed network than state-of-the-art methods. The performance of the DOA estimation using 1-bit ADC is improved to exceed that using 2-bit ADC.
\end{abstract}


\IEEEpeerreviewmaketitle

\section{Introduction}

\IEEEPARstart{E}{xtending} a massive multiple-input-multiple-output (MIMO) system \cite{risi2014massive} to the millimeter wave (MMV) band will bring higher hardware costs and attenuate the signal significantly, so intelligent reflecting surface (IRS) \cite{8937491,zheng2020intelligent} as a new method
 is introduced to enhance the wireless link and reduce costs of additional base stations. For the massive MIMO system, when the low-cost analog to digital converters (ADCs) are used in the IRS equipped with radio frequency (RF) chains, the ADC resolution is decreasing, and the signals will be deteriorated severely by the quantization noise. It is expected to ensure the quality of signal processing while decreasing costs. 

Direction of angle (DOA) estimation with 1-bit quantized measurements has been studied for many years. Ref. \cite{1039405} considers the influence of 1-bit measurements on the estimation accuracy in the scenario with additional white Gaussian noise (AWGN), and the estimation for the covariance matrix is also investigated. Similar to the DOA estimation, a line spectrum estimation with quantized signals is investigated in ref. \cite{824661}, where design rules for antialiasing filter and sampling rate are proposed. They are applied to delta-sigma ADC to reduce the quantization noise, which has been widely used in the industry. Additionally, 1-bit quantized sparse arrays proposed in \cite{7952732} perform better than the unquantized uniform linear arrays (ULAs) in the DOA estimation problem.

Deep learning has aroused great interest in the past years and has been combined with wireless communication and radar signal processing. In \cite{8400482}, a channel and DOA estimation method based on deep neural network (DNN) is proposed in the massive MIMO system. Usually, for the DOA estimation using DNN, the output of DNN is the spatial angle, and this scheme can achieve high resolution. Moreover, there are some other kinds of DNN structures for the DOA estimation, such as the classifier-based network \cite{8485631}, which divides the spatial angle into grids. The input of the network can also use a covariance matrix instead of the raw baseband signals. Denoising convolutional neural network (DnCNN) is first introduced in \cite{7839189} to remove Gaussian noise in image, and ref. \cite{8822737} applies DnCNN in the array signal processing with a better performance than the atomic norm minimization (ANM) denoising method. 

However, for the quantized signals, in addition to the reconstructed covariance matrix in \cite{1039405}, there are few ways to deal with the reconstruction of quantized signals. In the hardware, the sigma-delta ADC can decrease the quantization noise \cite{824661}, but it is expensive when the sampling rate increases, which is not applicable in the massive MIMO system and IRS who possess large-scale array.

In this paper, we propose a reconstruction method for quantized signal based on DNN, which is composed of fully connected (FC) layers, batch normalization (BN)  unit, residual block, and rectified linear unit (ReLU) activation function. Through offline training, ADC precision can be improved without additional performance loss. Then, the recovery performance for the quantized signal is evaluated by measuring the accuracy of the DOA estimation. Moreover, the reconstruction performance of the proposed network is also compared with state-of-the-art methods.

The paper is structured as follows. The signal and quantization noise model are presented in section II. In Section III, we propose the scheme for signal reconstruction. Simulation results and discussion are illustrated in Section IV. Finally, in Section V, the paper is concluded.

\begin{figure*} \centering{\includegraphics[width=\textwidth]{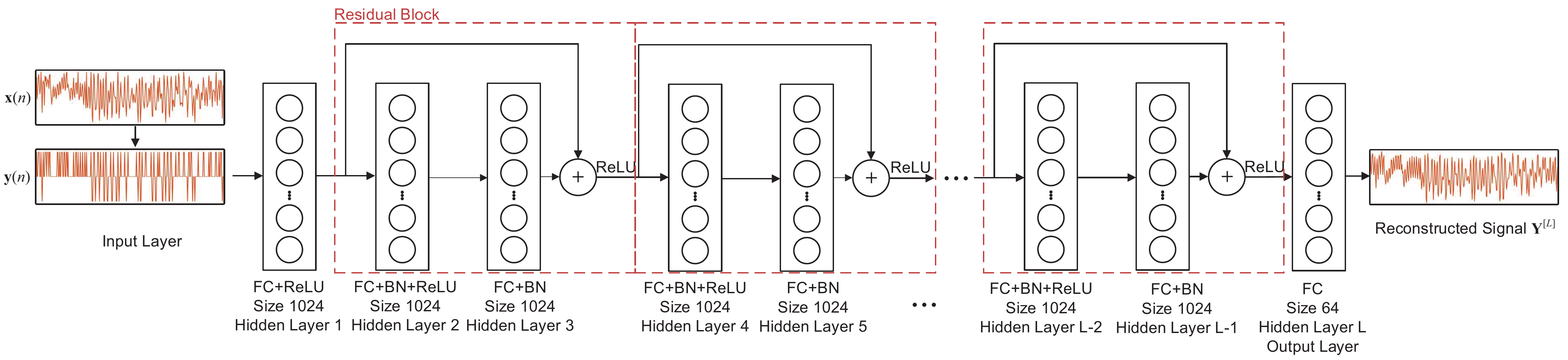}} \caption{The proposed DNN structure.} 
\end{figure*}

\section{Signal and Quantization Noise Model}
Considering a ULA with $M$ sensors, $d$ is array  spacing, and $\lambda$ is wavelength of signals. This ULA array is used to estimate the DOA of the $K$ far-field signals. For the $k$-th signal $\left(k=1,2,3,…,K\right)$, the complex signal and the DOA are denoted as $s_{k}$ and $\theta_{k}$, respectively. During the $n$-th sampling time, when the low-cost ADC is used, the received signal in ULA can be expressed as a baseband signal 
\begin{equation}   
\begin{aligned}  
\mathbf{y}(n)=\sum_{k=1}^{K}s_{k}(n)\mathbf{a}(\theta_{k})+\mathbf{e}(n)+\mathbf{q}(n)=\mathbf{x}(n)+\mathbf{q}(n),
\end{aligned} 
\end{equation}
where the received signal from $M$ sensors is $\mathbf{x}(n)=\sum_{k=1}^{K}s_{k}(n)\mathbf{a}(\theta_{k})+\mathbf{e}(n)\in{\mathbb{C}^{M\times1}}$, and the steering vector is $\mathbf{a}(\theta_{k})=\left[1,e^{j2\pi\frac{d}{\lambda}\sin(\theta_{k})},...,e^{j2\pi\frac{d(M-1)}{\lambda}\sin(\theta_{k})}\right]^{\text{T}}$. The quantized signal is $\mathbf{y}(n)\in{\mathbb{C}^{M\times1}}$, the complex additive Gaussian noise vector is $\mathbf{e}(n)=\left[e_{1}(n),...,e_{M}(n)\right]^{\text{T}}$, and $\mathbf{q}(n)=\left[q_{1}(n),...,q_{M}(n)\right]^{\text{T}}$ denotes the quantization noise. The $m$-th entry in $\mathbf{q}(n)$ is
\begin{equation}   
\begin{aligned}  
   q_{m}(n)=\text{round}&\left(\frac{x_{m}(n)}{\Delta(B)}\right)\Delta(B)-x_{m}(n), \\&m=1,...,M,
\end{aligned}  
\end{equation}
where round($\cdot$) is round operation,     $\Delta(B)=2V/2^{B}$, and $V$ is maximum input voltage of ADC. The real part and imaginary part of $x_{m}(n)$ are distributed between $–V$ and $V$. $q_{m}$ is quantization noise produced by $B$-bit ADC, and follows an uniform distribution from $-0.5\Delta(B)$ to $0.5\Delta(B)$. We use rounding quantization, which resulting in that we have $2^{B}+1$ quantization level for $B$ bits quantization. All the quantizations used below are the same as those used in this section. In this paper, we will estimate the DOA $\theta_{k}$ from the quantized signals $\mathbf{y}(n)$.
\begin{figure*}   \centering   \subfigure[Fine-tuned DNN with different layers.]{\includegraphics[width=2.22in]{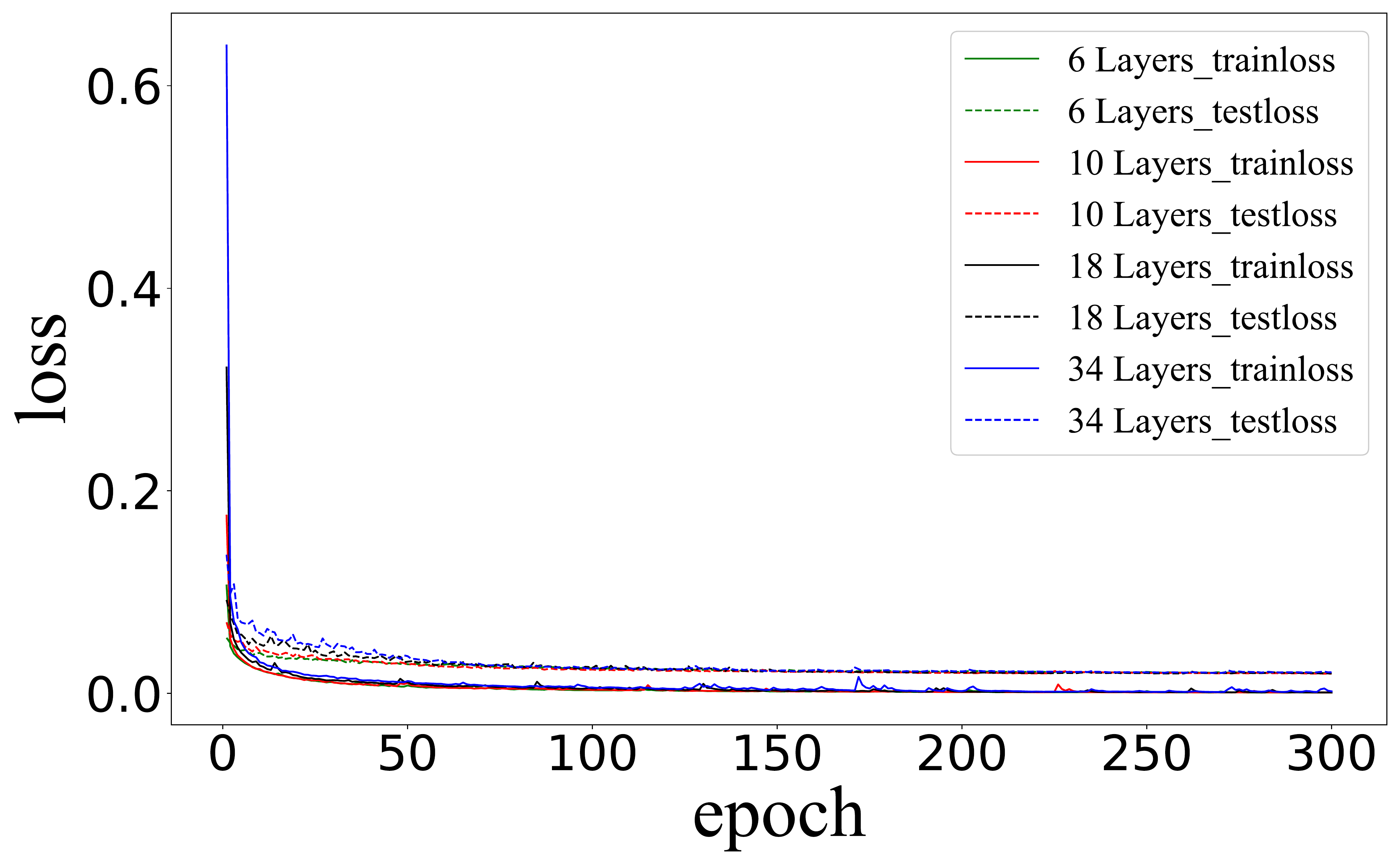}}   \subfigure[Fine-tuned DNN with different neurons.]{\includegraphics[width=2.22in]{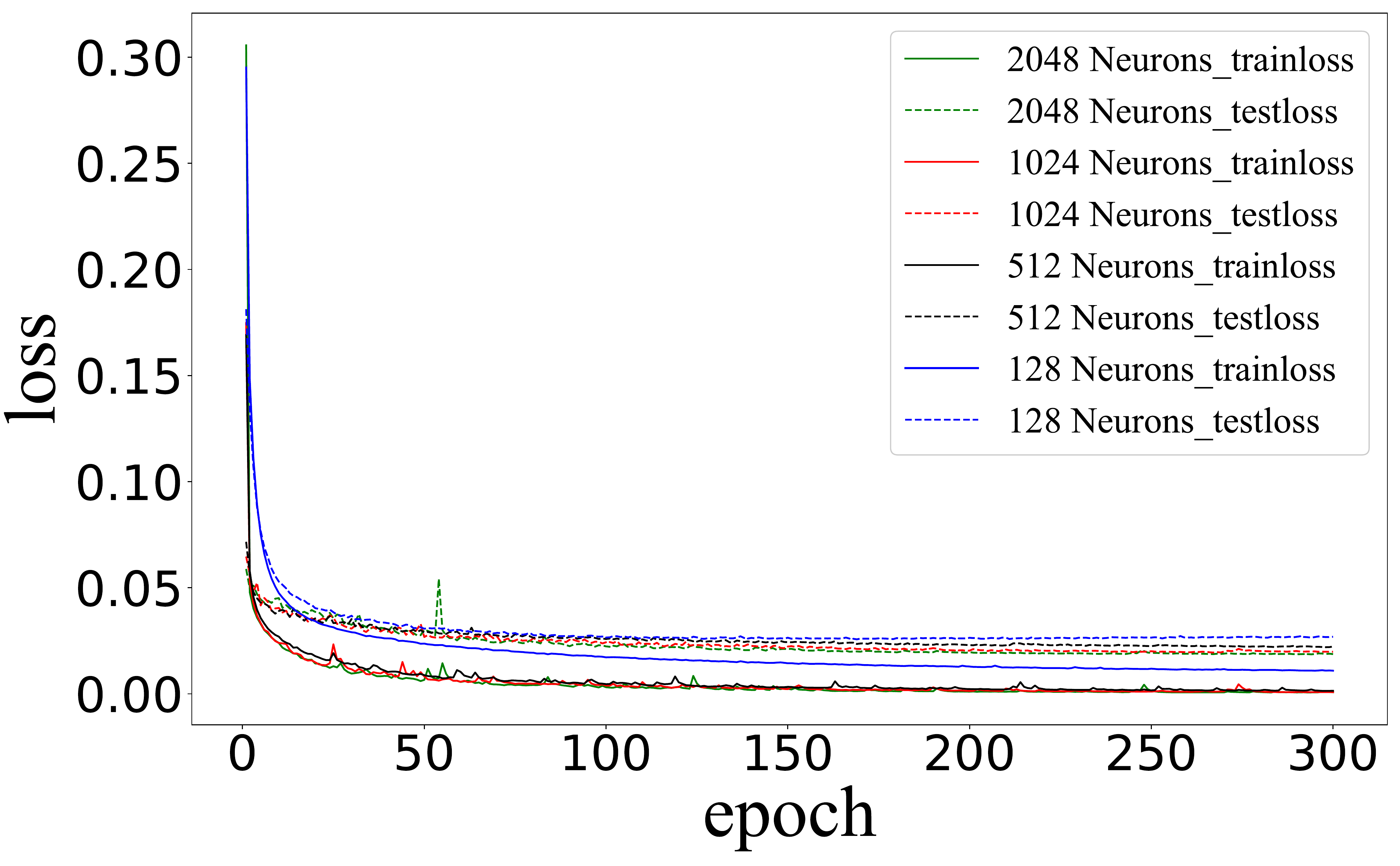}}      \subfigure[Fine-tuned DNN with different structure.]{\includegraphics[width=2.22in]{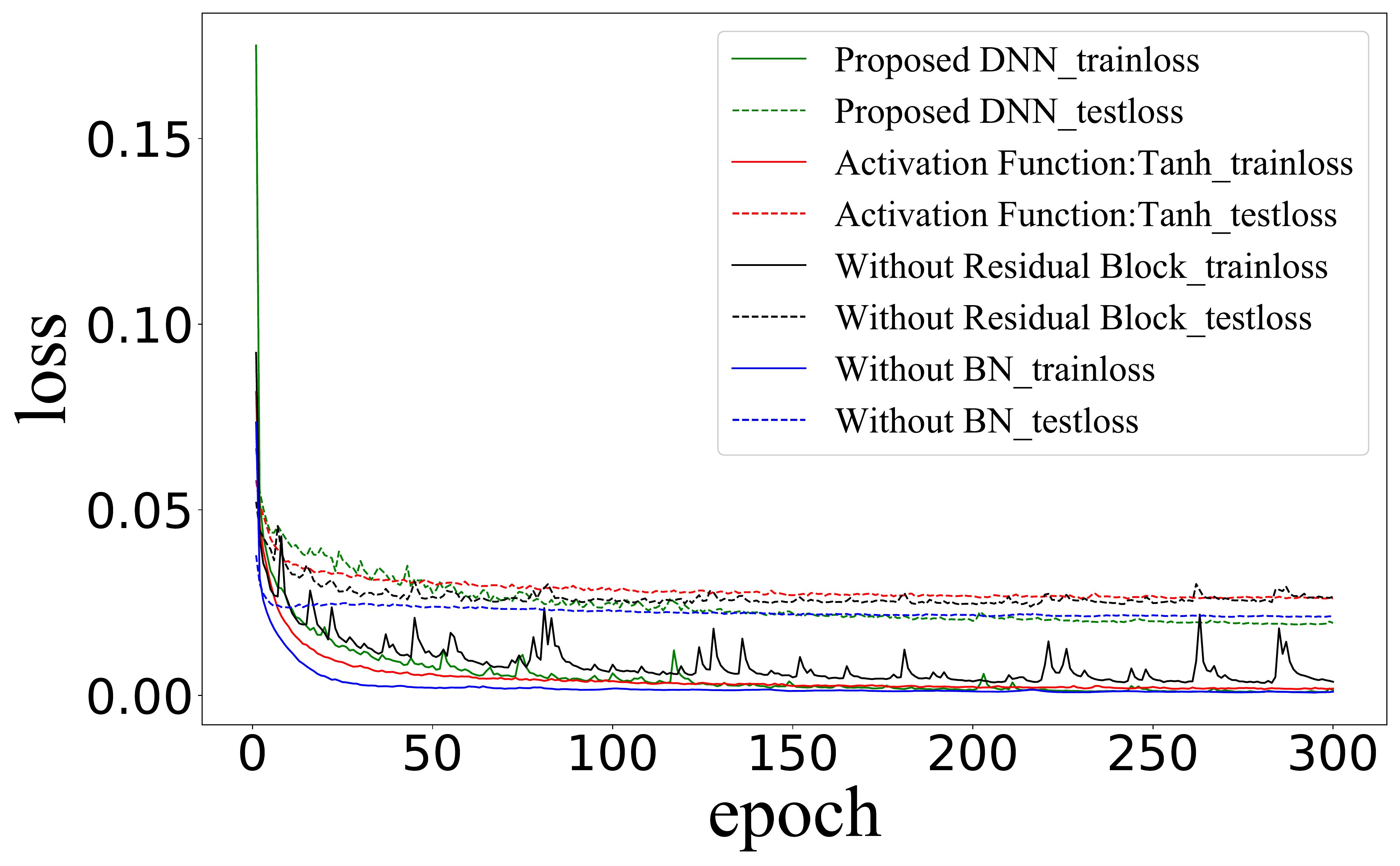}}     \caption{Comparison of different networks in 300 iterations.}  \end{figure*}

\section{Proposed Scheme For Signal Reconstruction}
\label{sec:guidelines}
To recover the information from the quantized signals, we propose a DNN network with FC layers. With the quantized signal as input, the network is trained to output the original signal  (unquantized signal, $\mathbf{x}(n)$) .

As shown in Fig. 1, the DNN has $L$ hidden layers, we employ the ReLU \cite{nair2010rectified} as our activation function, which can be given by
\begin{equation}        
	f_{\text{ReLU}}(x)=\max(0,x).
\end{equation}
We also employ the residual block \cite{he2016deep} to prevent overfitting, gradient explosion and disappearance. The real and  imaginary parts of $\mathbf{y}(n)$ are separated and reshaped into a long vector. For $M$ elements in the ULA, each signal being input to DNN is a tensor $\mathbf{Y}^{[0]}\in{\mathbb{R}^{1\times{2M}}}$. The first layer of this network is FC layer, and ReLU is applied to the output of first FC layer, so the output of first hidden layer can be expressed as
\begin{equation}        
	\mathbf{Y}^{[1]}=f_{\text{ReLU}}\left(\mathbf{Y}^{[0]}\mathbf{W}^{[1]}\right),
\end{equation}
where $\mathbf{W}^{[1]}\in{\mathbb{R}^{2M\times{N_{1}}}}$ is the weight of the first FC layer, $\mathbf{b}^{[1]}\in{\mathbb{R}^{N_{1}}}$ is its bias term, and $N_{l}$ is the amount of neuron in hidden layer $l$.

Residual block is applied to layers after the hidden layer 1. For the first residual block, the following calculation is performed 
\begin{equation}   
\begin{aligned}        	
 	&\mathbf{Y}^{[2]}=f_{\text{ReLU}}\left(f_{\text{BN}}\left(\mathbf{Y}^{[1]}\mathbf{W}^{[2]}+\mathbf{b}^{[2]}\right)\right),\\
 	&\mathbf{Y}^{[3]}=f_{\text{ReLU}}\left(\mathbf{Y}^{[1]}+f_{\text{BN}}\left(\mathbf{Y}^{[2]}\mathbf{W}^{[3]}+\mathbf{b}^{[3]}\right)\right), 
\end{aligned}  
\end{equation}
where $\mathbf{W}^{[l]}\in{\mathbb{R}^{N_{l-1}\times{N_{l}}}}, \mathbf{b}^{[l]}\in{\mathbb{R}^{N_{l}}}$ for $l=2,3,...,L$. In residual block, we applied BN \cite{ioffe2015batch}  to the output of each FC layer. BN uses the mean and standard deviation on small batches to continuously adjust the intermediate output of the neural network during training, thereby making the intermediate output of the entire neural network more stable. BN can be expressed as
\begin{equation}   
\begin{aligned}   
f_{\text{BN}}\left(\mathbf{Y}^{[l]}\right)=\mathbf{\gamma}&\odot\frac{\mathbf{Y}^{[l]}-E\left(\mathbf{Y}^{[l]}\right)}{\sqrt{var\left(\mathbf{Y}^{[l]}\right)+\epsilon}}+\mathbf{\beta}, 
\\&l=2,...,L-1,
\end{aligned}  
\end{equation} 
where $\epsilon$ is a small constant to ensure denominator is greater than 0. $\mathbb{E(\cdot)}, var(\cdot)$ calculate expectation and variance. $\mathbf{\gamma}, \mathbf{\beta}$ are scaling and shift parameters, respectively. $\odot$ denotes the element-wise product.

To obtain the reconstruction of quantized signals, the activation function is not used in the last layer, and only linear operations are performed
\begin{equation}    
	\mathbf{Y}^{[L]}=\mathbf{Y}^{[L-1]}\mathbf{W}^{[L]}+\mathbf{b}^{[L]}.    	 
\end{equation}

The loss function used for back propagation, reconstruction performance evaluation is as follows
\begin{equation}     
	f_{\text{loss}}\left(\mathbf{Y}^{[L]}\right)=\frac{1}{2M}\parallel \mathbf{Y}^{[L]}-\mathbf{\psi}(n)\parallel^{2}_{2},
\end{equation}
where $\mathbf{\psi}(n)=\left[\mathcal{R}\{\mathbf{y}(n)-\mathbf{q}(n)\}^{\text{T}},\mathcal{I}\{\mathbf{y}(n)-\mathbf{q}(n)\}^{\text{T}}\right]$, $\mathcal{R}\left\{\cdot\right\},\mathcal{I}\left\{\cdot\right\}$ take the real and imaginary parts of the complex vector, respectively. Under normal circumstances, multiple signals will be concatenated and input into the network, the loss value is the average for these signals.
\begin{figure*}   	\centering   	\subfigure[Performance of two method on small training set.]{\includegraphics[width=1.62in]{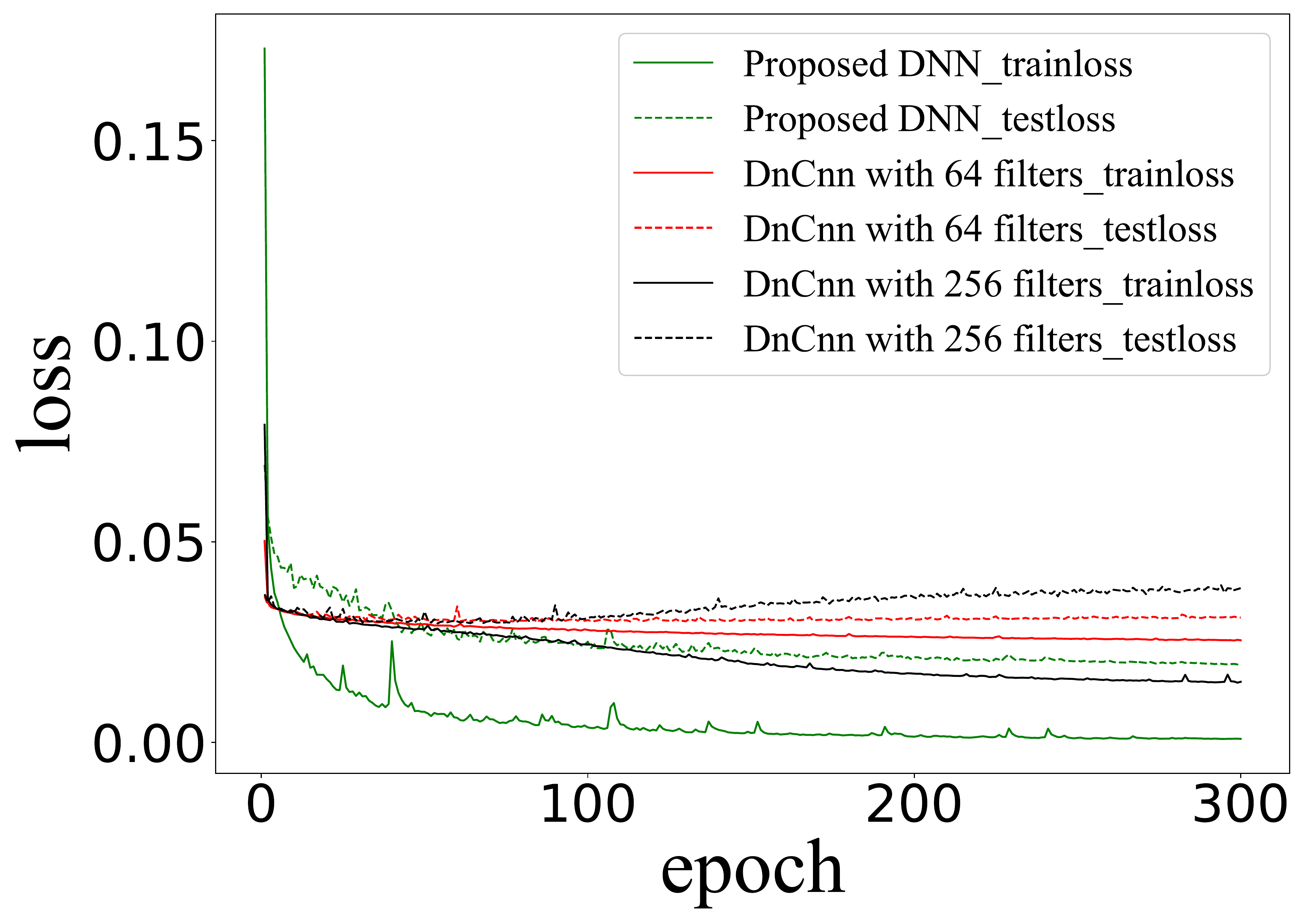}}   \subfigure[Test loss of proposed DNN in different SNR.]{\includegraphics[width=1.62in]{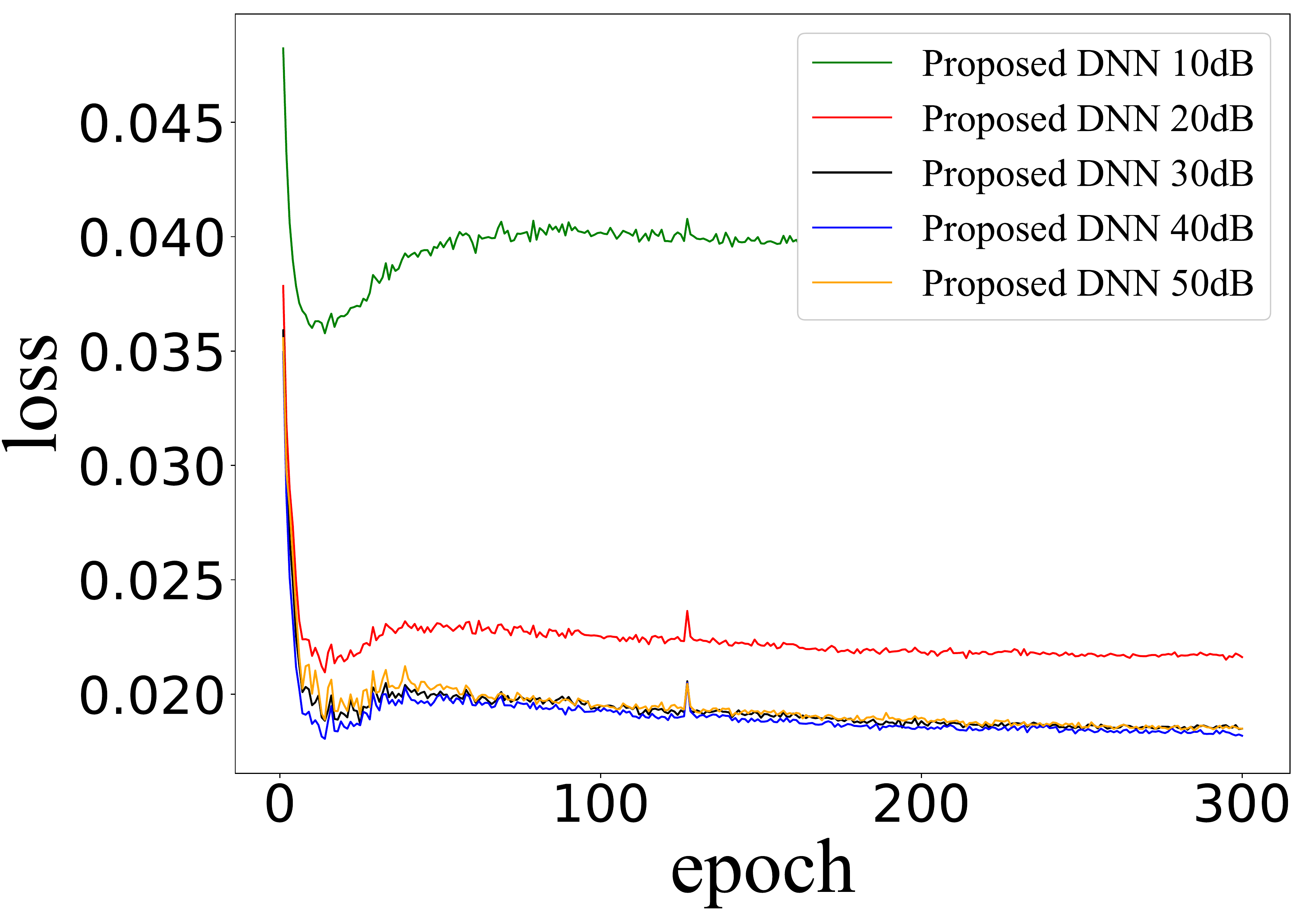}}      \subfigure[Test loss of DnCNN in different SNR.]{\includegraphics[width=1.62in]{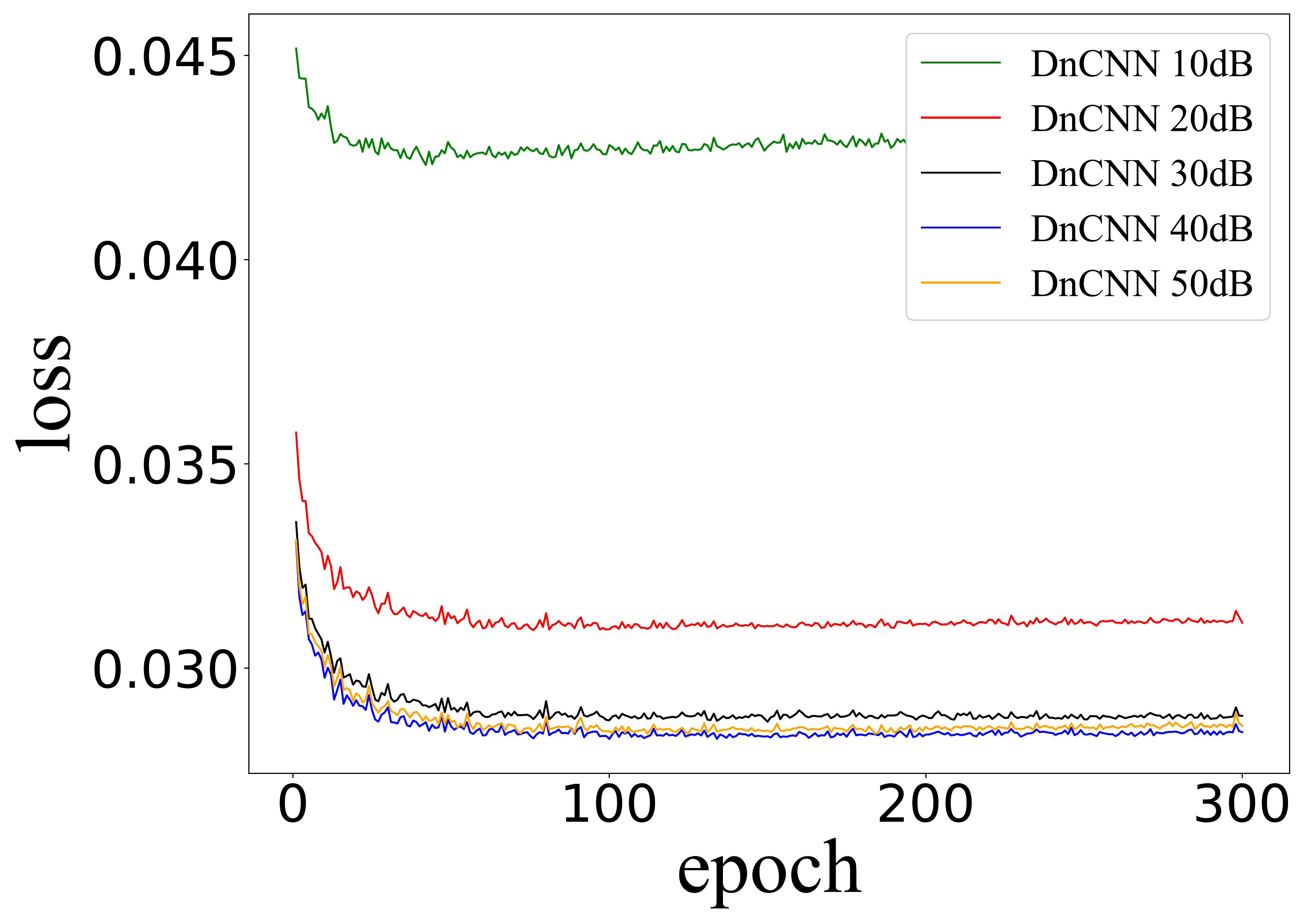}}      	\subfigure[MSE of two methods via MUSIC algorithm.]{\includegraphics[width=1.62in]{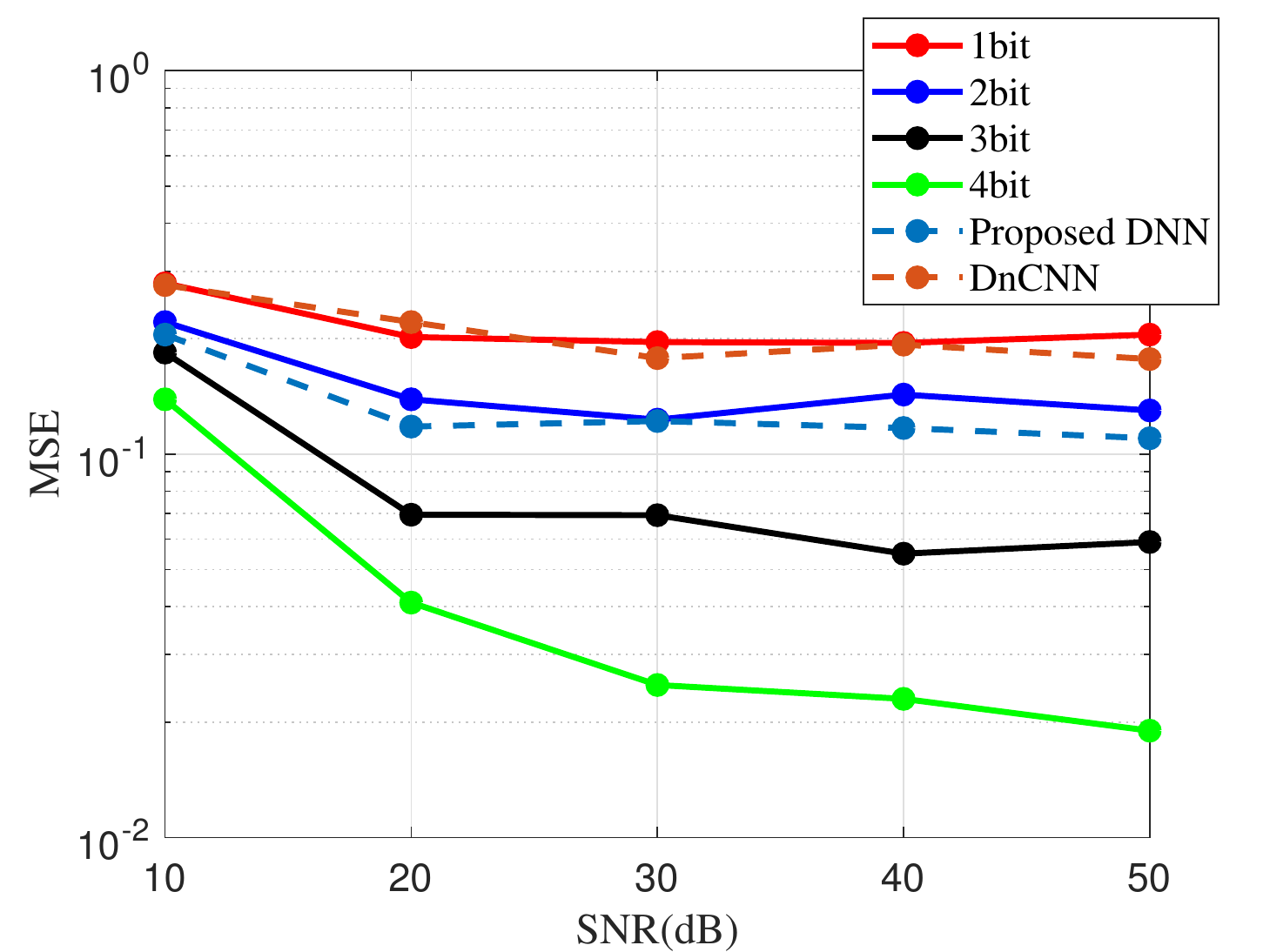}}	 	\caption{Proposed DNN vs DnCNN.}   \end{figure*}
\section{Simulation Results And Discussion}
A 32 elements ULA with half wavelength spacing between sensors is used to reconstruct signals containing 3 different spatial angles, where the spatial scope is $\left[\ang{-30},\ang{30}\right]$. Pytorch framework and Adam optimizer are used to implement proposed scheme. Learning rate is set to $10^{-3}$ without decay. In the meantime, we set $L$ to 10, $N_{l}$ $(l = 1,2,3,…,L-1)$ to 1024, $N_{L}$ to 64 and size of mini-batch to 256, and the data set is shuffled randomly. The numerical results in this section are generated by inputting 1-bit quantized signal into the proposed DNN. The experimental platform is a PC with graphics card NVIDIA 2080Ti. Acquiring codes at \url{https://github.com/hwfhwf/DNN-For-Signal-Reconstruction}.

First, by fine-tuning the proposed DNN to generate different networks, thus explaining why this network is used. Subsequently, comparing the DnCNN in \cite{8822737} and proposed DNN, multiple signal classification \cite{1143830} (MUSIC) algorithm is applied to evaluate neural network performance. The MUSIC algorithm uses 5 snapshots to compute the covariance matrix, and 1000 trials are performed. Finally, a method of compressing the model are examined, it can be deployed on mobile devices to reduce costs through numerical results.
\subsection{Assessment For Fine-Tuned Networks }

\begin{table}
	\caption{Training time required for networks with different numbers of neurons (300 iterations)} 
	\label{table} 
	\small
	\centering
	\normalsize
	\setlength{\tabcolsep}{3pt}
	\begin{tabular}{p{100pt}p{100pt}}
		\hline
		\multicolumn{1}{c}{$N_{l} (l=1,2,…,L-1)$} & \multicolumn{1}{c}{Training time (\emph{s})} \\ \hline
		\multicolumn{1}{c}{128}& \multicolumn{1}{c}{133.69}\\ 
		\multicolumn{1}{c}{512}& \multicolumn{1}{c}{134.73}\\ 
		\multicolumn{1}{c}{1024}& \multicolumn{1}{c}{169.59}\\ 
		\multicolumn{1}{c}{2048}& \multicolumn{1}{c}{338.14}\\ \hline
	\end{tabular}
\end{table}

The performance of different networks based on proposed DNN are tested in 300 iterations, with 20000 training signals and 1000 test signals for SNR = 50 dB. As Fig. 2(a) shown, it is clear that adding layers will not bring large benefits by changing the number of layers of proposed DNN, and will lead to gradient explosion. As depicted in Fig. 2(b), increasing the number of neurons $N_{l}$ brings obvious benefits on test set, but DNN with 2048 neurons will cost much more time in training, as shown in Table \uppercase\expandafter{\romannumeral1}. In Fig. 2(c),  proposed DNN is adjusted on activation function, with/without residual block, with/without BN, respectively. It is worth noting that even in the case where the network is not that deep, the residual block can still suppress the gradient explosion. Although the curve gap is not obvious during training, on the test set, the loss of proposed method is significantly smaller than networks that has been changed, implying that this network  performs better.

\subsection{Comparison Between Proposed DNN With DnCNN}

Fig. 3(a) compares the proposed DNN with DnCNN in \cite{8822737}. For DnCNN, the number of filters is adjusted to 64 and 256 in each layer, which can be seen that over-fitting occurs in DnCNN. Increasing the training set size to $10^5$, test set size to 5000 with different $\text{SNR}\in{\{10\text{ dB},20\text{ dB},30\text{ dB},40\text{ dB},50\text{ dB}\}}$, the data set is evenly distributed in each SNR. Fig. 3(b) and Fig. 3(c) contain test loss of two network in different SNR. As the size of the training set increases, over-fitting no longer occurs for DnCNN, consequently more data is required. From two figures, the proposed DNN possesses better reconstruction capability, due to lower test loss in the whole SNR range. As shown in Fig. 3(d), the reconstructed signals output by the network corresponding to Fig.3(b) and Fig. 3(c) are used to estimate the DOA via MUSIC algorithm. As comparison, the DOA of 1$\thicksim$4 bits quantized signals also been estimated to evaluate signal reconstruction level. For estimated angle $\hat{\theta}_{k}$, MSE of angle can be expressed as: $\text{MSE}=\frac{1}{K}\sum_{k=1}^{K}\left(\hat{\theta}_{k}-\theta_{k}\right)^{2}$. With same quantized signal, reconstruction effect of DnCNN is very weak. Through proposed DNN, the performance of the DOA estimation using 1-bit ADC is improved to exceed that using 2-bit ADC.

\subsection{Further Training For Proposed DNN}
\begin{figure}     	
	\centering
	\subfigure[Signal reconstructed by Proposed DNN comparing with different quantized signal.]{\includegraphics[width=1.62in]{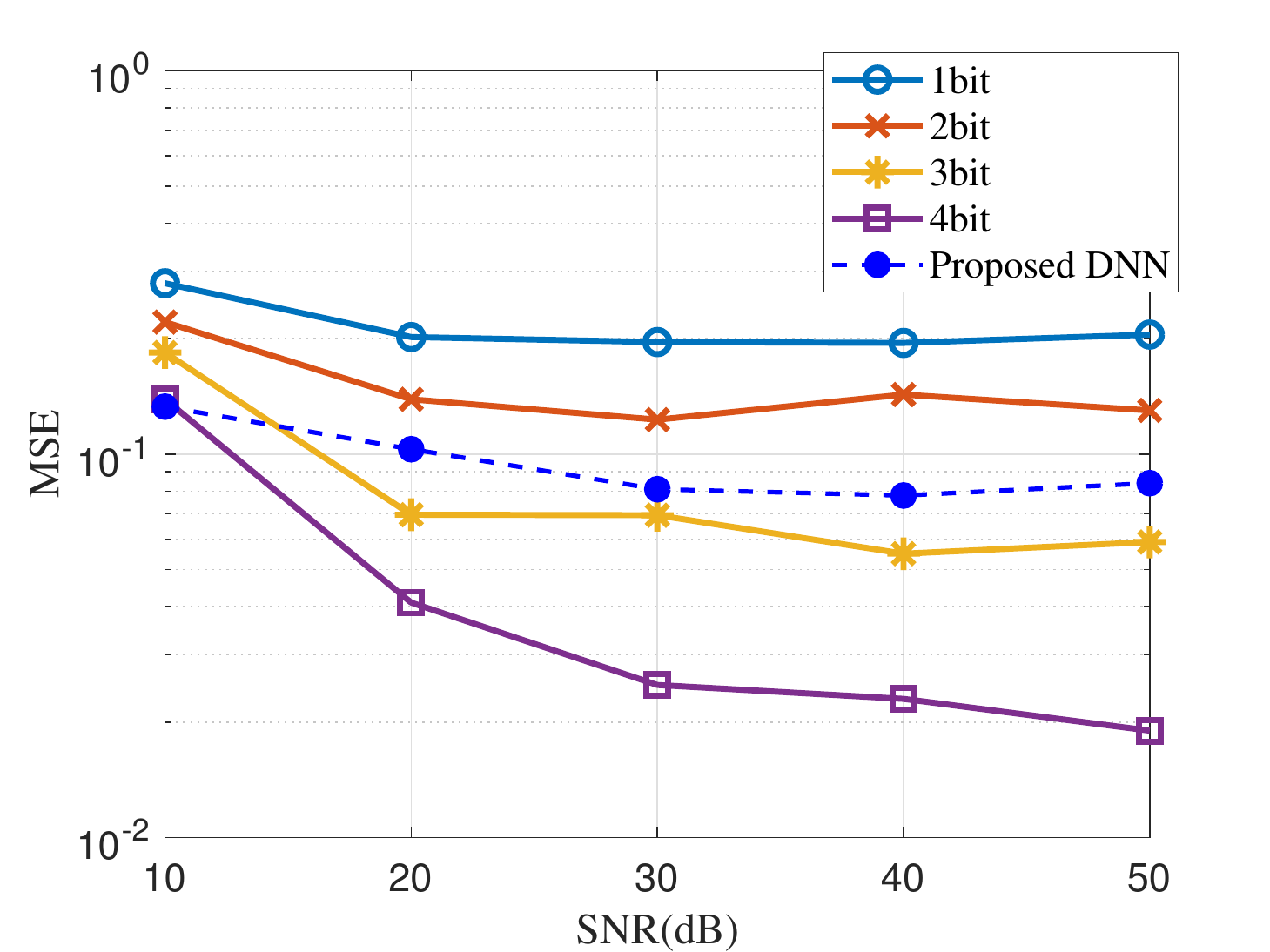}}   \subfigure[DOA estimation with reconstructed signal, 2 bit and 3 bit quantized signal.]{\includegraphics[width=1.62in]{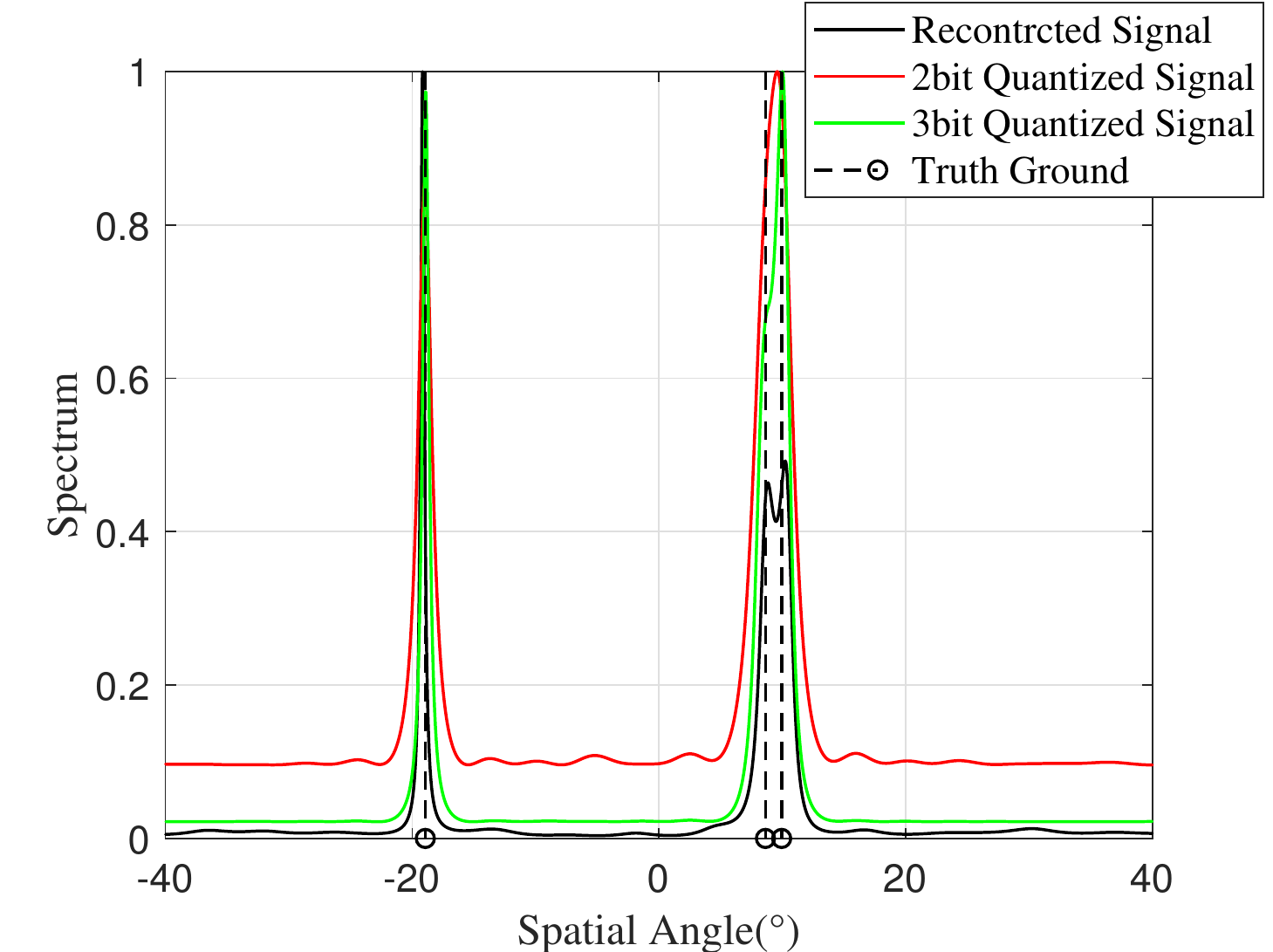}}  
	\caption{The reconstruction effect of proposed DNN in large training set.}  
 \end{figure}

Increasing the size of training set to $10^7$ for each SNR to make the network more generalized. After 300 iterations, 1-bit quantized signal are input into the trained network to generate  reconstructed signal, and the MUSIC algorithm is used to verify the recovery effect. By comparing with the MSE of 1$\thicksim$4-bit quantized signal, estimate the improvement effect obtained by this method. As is shown in Fig. 4(a), owing to proposed DNN, the performance of the DOA estimation using 1-bit ADC is improved to exceed that using 2-bit ADC on each SNR. In Fig. 4(b), the true angles of the signal are $\ang{-18.9346},\ang{8.6346},\ang{9.9462}$. In the scenario of SNR = 50 dB, the signal reconstructed by proposed network can achieve better resolution than 3-bit quantized signal around $\ang{9}$.

\subsection{Compressed Model}
\begin{table}
	\caption{MSE comparison between the simplified and the original model}  	
	\label{table}  	
	\centering 	
	\normalsize 
	\setlength{\tabcolsep}{3pt}
	\begin{tabular}{ccccccccccc} 
		\hline
		SNR (dB)  & 
		\multicolumn{2}{c}{30} & 
		\multicolumn{2}{c}{40} & 
		\multicolumn{2}{c}{50} \\\hline
		 Precision & 16 FP      & 32 FP     & 16 FP      & 32 FP     & 16 FP      & 32 FP     \\ MSE       & 0.081      & 0.081     & 0.080      & 0.078     & 0.081      & 0.084   \\\hline
	\end{tabular} 
\end{table}
For field programmable gate array (FPGA) or embedded devices, the network model must be compressed as much as possible to ensure high processing speed, at the same time the accuracy of signal processing must be guaranteed. There are some methods to compress the model such as precision reduction and pruning for parameters in neural network. Our network parameters are initially single-precision floating-point (32 FP) numbers, turning them to half-precision floating-point (16 FP), then we can get 50\% reduction in model size and faster data processing speed. From Table \uppercase\expandafter{\romannumeral2}, precision from 32 FP to 16 FP hardly changes the MSE, it is feasible that continue to decrease the precision of parameters to achieve speed-performance trade-off. The operation we used above are for the trained network parameters. Greater benefits can be achieved by applying them to the training process, which is deeply discussed in \cite{han2015deep}, including network quantization, pruning, weight sharing et al.

\section{Conclusion}

 This paper has proposed a DNN to decrease quantization noise and reconstruct the quantized signal. The numerical results have shown that under sufficient training, through proposed DNN, the performance of the DOA estimation using 1-bit ADC is improved to exceed that using 2-bit ADC, consequently ADCs with lower resolution can be employed without performance loss. This method can be applied to massive MIMO system, IRS with RF chains, thus greatly reducing costs.

\bibliographystyle{unsrt} 
\bibliography{ref} 

\end{document}